\documentclass[%
  twoside,
  reprint,
  amsmath,amssymb,
  aps,
  pra,
  nofootinbib,
  showpacs,
  superscriptaddress,
  a4paper
]{revtex4-1}

\usepackage{graphicx}%
\usepackage[usenames,dvipsnames]{xcolor}
\usepackage{siunitx}
\usepackage{subfigure}

\usepackage{lipsum}

\usepackage[
text={7.3in,10in},centering,
total={6.5in,8.75in}, top=1in, left=0.52in, includefoot,
]{geometry}

\usepackage[
  bookmarks=true,
  colorlinks,
  linkcolor=blue,
  urlcolor=blue,
  citecolor=blue,
  plainpages=false,
  pdfpagelabels,
  final,
  breaklinks=true
]{hyperref}
\hypersetup{
pdftitle={High harmonic interferometry of the Lorentz force in strong mid-infrared laser fields}, 
pdfauthor={Emilio Pisanty, Daniel D. Hickstein, Benjamin R. Galloway, Charles G. Durfee, Henry C. Kapteyn, Margaret M. Murnane and Misha Ivanov}
}

\usepackage{natbib}
\makeatletter \def\NAT@def@citea{\def\@citea{\NAT@separator\,}} \makeatother
\newcommand{\citer}[1]{Ref.~\citealp{#1}}

\newcommand{\reffig}[1]{Fig.~\ref{#1}}

\renewcommand{\d}{\ensuremath{\textrm{d}}}

\newcommand{\ket}[1]{\ensuremath{\left|#1\right\rangle}}

\newcommand{\vb}[1]{\mathbf{#1}}
  \newcommand{\vbv}{\vb{v}}
  \newcommand{\vbr}{\vb{r}}
  \newcommand{\vba}{\vb{A}}
  \newcommand{\vbf}{\vb{F}}
  \newcommand{\vbb}{\vb{B}}
  \newcommand{\vbp}{\vb{p}}
  \newcommand{\vbk}{\vb{k}}
  \newcommand{\vbd}{\vb{d}}

\newcommand{\gradA}{\nabla\!\vba}
\newcommand{\vbpi}{\boldsymbol{\pi}}

\newcommand{\volkov}[2]{\ket{\Psi^\mathrm{(V)}_{#1}(#2)}}

\newcommand{\tn}{{t_0}}
\newcommand{\cc}{\ensuremath{\mathrm{c.c.}}}
\newcommand{\eps}{\varepsilon}

\begin{document}

\title{High harmonic interferometry of the Lorentz force in strong mid-infrared laser fields}

\author{Emilio Pisanty}
 \email{e.pisanty11@imperial.ac.uk}
 \affiliation{Blackett Laboratory, Imperial College London, South Kensington Campus, SW7 2AZ London, United Kingdom}
 \affiliation{Max Born Institute for Nonlinear Optics and Short Pulse Spectroscopy, Max Born Strasse 2a, 12489 Berlin, Germany}
\author{Daniel D. Hickstein}
 \email{danhickstein@gmail.com}
 \affiliation{JILA -- Department of Physics, University of Colorado and NIST, Boulder, Colorado 80309, USA}
\author{Benjamin R. Galloway}
 \affiliation{JILA -- Department of Physics, University of Colorado and NIST, Boulder, Colorado 80309, USA}
\author{Charles G. Durfee}
 \affiliation{JILA -- Department of Physics, University of Colorado and NIST, Boulder, Colorado 80309, USA}
 \affiliation{Colorado School of Mines, Department of Physics, Golden, Colorado 80401 USA}
\author{Henry C. Kapteyn}
\author{Margaret M. Murnane}
 \affiliation{JILA -- Department of Physics, University of Colorado and NIST, Boulder, Colorado 80309, USA}
\author{Misha Ivanov,}
 \email{m.ivanov@imperial.ac.uk}
 \affiliation{Blackett Laboratory, Imperial College London, South Kensington Campus, SW7 2AZ London, United Kingdom}
 \affiliation{Max Born Institute for Nonlinear Optics and Short Pulse Spectroscopy, Max Born Strasse 2a, 12489 Berlin, Germany}
 \affiliation{Department of Physics, Humboldt University, Newtonstrasse 15, 12489 Berlin, Germany}

\date{\today}

\begin{abstract}
The interaction of intense mid-infrared laser fields with atoms and molecules  leads to a range of new opportunities, from the production of bright, coherent radiation in the soft x-ray range to imaging molecular structures and  dynamics with attosecond temporal and sub-angstrom spatial resolution. However, all these effects, which rely on laser-driven recollision of an electron removed by the strong laser field and the parent ion, suffer from the rapidly increasing role of the magnetic field component of the driving pulse: the associated Lorentz force pushes the electrons off course in their excursion and suppresses all recollision-based processes, including high harmonic generation, elastic and inelastic scattering. Here we show how the use of two non-collinear beams with opposite circular polarizations produces a forwards ellipticity which can be used to monitor, control, and cancel the effect of the Lorentz force. This arrangement can thus be used to re-enable recollision-based phenomena in regimes beyond the {\it long}-wavelength breakdown of the dipole approximation, and it can be used to observe this breakdown in high-harmonic generation using currently-available light sources.
\end{abstract}

\maketitle

Strong-field phenomena benefit from the use of long-wavelength drivers. Indeed, for sufficiently intense fields, the energy of the interaction scales as the square of the driving wavelength, since with a longer period the electron has more time to harvest energy from the field. In particular, long-wavelength drivers allow one to extend the generation of high-order harmonics~\cite{corkum_hhg-review_2007, HHGTutorial, kohler_chapter_2012} towards the production of short, bright pulses of x-ray radiation, currently reaching  into the $\SI{}{keV}$ range with thousands of harmonic orders~\cite{popmintchev_record_2012}, and with driving laser wavelengths as long as $\SI{9}{\micro\metre}$ under consideration~\cite{hernandez_nine-micron_2013,zhu_non-dipole_2016}.

However, this programme runs into a surprising limitation in that the dipole approximation breaks down in the \textit{long} wavelength regime: as the wavelength increases, the electron has progressively longer times to accelerate in the field, and the magnetic Lorentz force $\vbf_\mathrm{m}=\vbv/c \times \vbb$ becomes significant~\cite{reiss_dipole-approximation_2000}. This pushes the electron along the laser propagation direction 
and, when strong enough, makes
the electron wavepacket completely miss its parent ion, quenching all recollision phenomena, including in particular high harmonic generation~\cite{potvliege_photon_2000, walser_hhg_2000,kylstra_photon_2001, kylstra_photon_2002, chirila_analysis_2004, milosevic_relativistic_2002, milosevic_relativistic_2002-1, emelina_possibility_2014}.

Multiple schemes have been proposed to overcome this limitation, both on the side of the medium, from antisymmetric molecular orbitals~\cite{fischer_enhanced_2006} through relativistic beams of highly-charged ions~\cite{avetissian_high-order_2011} to exotic matter like positronium~\cite{hatsagortsyan_microscopic_2006} or muonic atoms~\cite{muller_exotic_2009}, and on the side of the driving fields, including counter-propagating mid-IR beams~\cite{taranukhin_relativistic_2000, verschl_relativistic_2007}, the use of auxiliary fields propagating in orthogonal directions~\cite{chirila_nondipole_2002}, fine tailoring of the driving pulses~\cite{klaiber_fully_2007}, counter-propagating trains of attosecond pulses~\cite{kohler_phase-matched_2011} in the presence of strong magnetic fields~\cite{verschl_refocussed_2007}, and collinear and non-collinear x-ray initiated HHG~\cite{klaiber_coherent_2008, kohler_macroscopic_2012}, though in general these methods tend to be challenging to implement. Perhaps most promisingly, one can also use the slight ellipticity in the propagation direction present in very tightly focused laser beams~\cite{lin_tight-focus_2006, galloway_lorentz_2016} and in waveguide geometries.

\begin{figure}[hb]
\smash{
  \subfigure{\label{fig:figureonea}}
  \subfigure{\label{fig:figureoneb}}
  \subfigure{\label{fig:figureonec}}
  \subfigure{\label{fig:figureoned}}
  \subfigure{\label{fig:figureonee}}
  \subfigure{\label{fig:figureonef}}
}
\vspace{-2mm}
\includegraphics[width=\columnwidth]{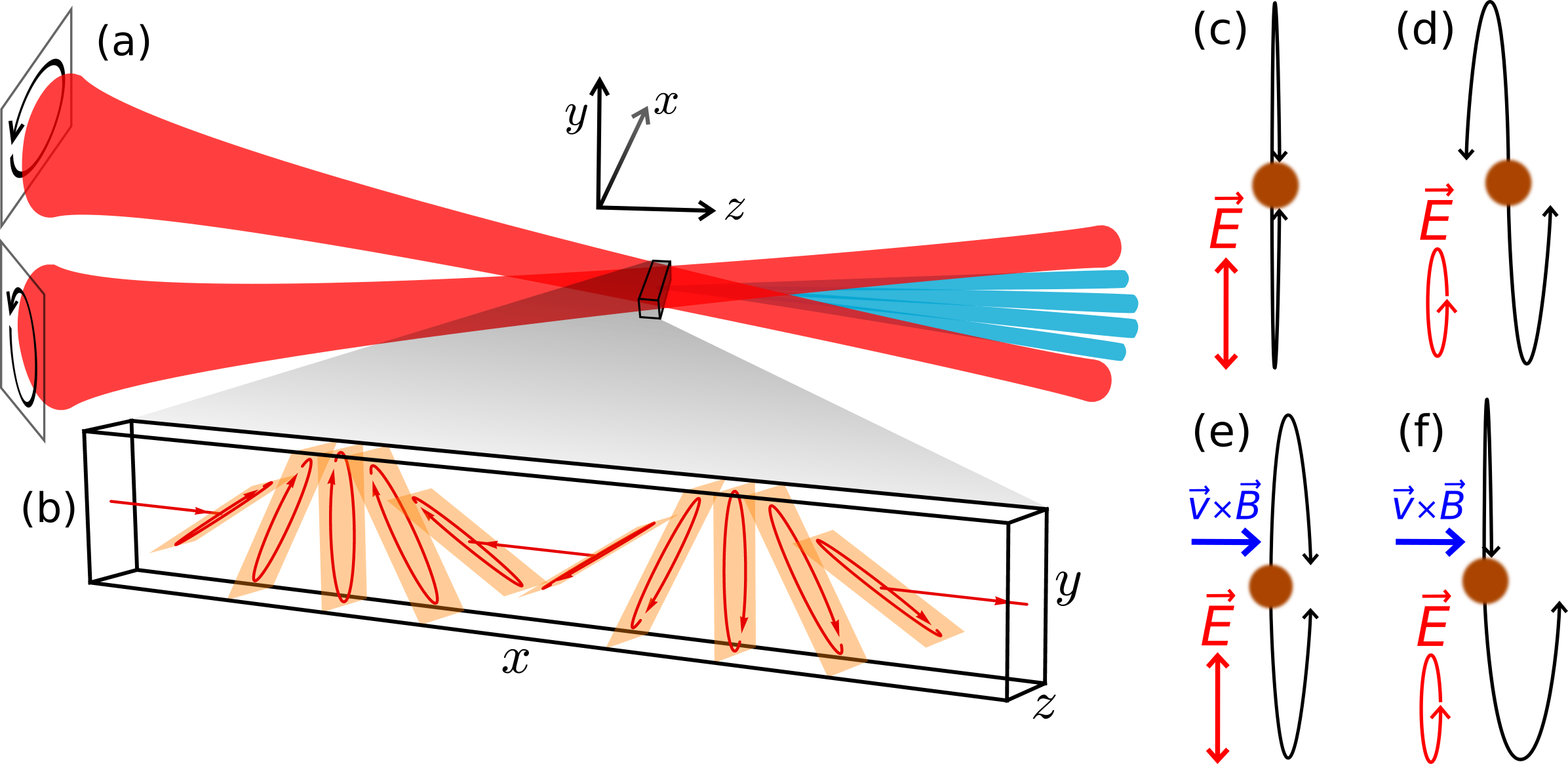} 
\caption{
Two non-collinear circularly polarized beams (a) produce a polarization gradient across their focus in the transverse direction (b), including points with forwards ellipticity. This takes the usual linear trajectories (c) on a path to miss the ion~(d), cancelling out the effect of the Lorentz force (e) to achieve one recollision per cycle~(f), re-enabling the harmonic emission.
}
\label{fig:figureone}
\end{figure}

Here we propose a simpler method for attaining a forwards ellipticity that can act in the same direction as the Lorentz force, thereby re-enabling the harmonic emission in the presence of magnetic effects, through the use of two non-collinear counter-rotating circularly polarized beams of equal intensity and wavelength, as shown in \reffig{fig:figureonea}. Normally, adding counter-rotating circular polarizations at the same frequency results in linear polarizations; for non-collinear beams, however, the planes of polarization do not quite match, and this means that at certain positions in the focus their components along the centerline will add constructively. This results in a forwards ellipticity: that is, elliptical polarizations with the unusual feature that the minor axis of the polarization ellipse is aligned along the centerline of the beam propagation.

This forwards ellipticity will tend to act in opposite directions for electrons released in each half-cycle (shown in \reffig{fig:figureoned}), as opposed to the Lorentz force, which acts always in the forwards direction (\reffig{fig:figureonee}), so one of the trajectories does return to the ion (\reffig{fig:figureonef}). One important consequence is that the symmetry between the two half-cycles is broken~\cite{galloway_lorentz_2016,averbukh_stability_2002}, creating an unbalanced interferometer. This allows one to clearly show the action of the magnetic Lorentz force on the continuum electron in HHG with high sensitivity, to complement the experimental confirmation of the long-wavelength breakdown of the dipole approximation in ionization experiments~\cite{smeenk_partitioning_2011, ludwig_breakdown_2014}.

Here we extend the description of nondipole HHG to cover non-collinear beam configurations. We show that non-collinear beams can indeed recover harmonic emission from damping by nondipole effects, and that the even harmonics, as a signature of the nondipole effects, are readily accessible to currently available laser sources.

The generation of harmonics using opposite circular polarizations, known as `bicircular' fields, has been the subject of theoretical study for some time~\cite{EichmannExperiment, SFALong, SFAMilosevicBecker, pisanty_spin-conservation_2014, milosevic_circularly_2015, medisauskas_generating_2016}, and it reached fruition with the use of an $\omega$-$2\omega$ collinear scheme to produce circularly polarized high harmonics~\cite{fleischer_spin_2014, kfir_generation_2015}. The use of non-collinear beams was demonstrated recently~\cite{hickstein_non-collinear_2015}, and it permits the angular separation of the circular harmonics, with opposite helicities appearing on opposite sides of the far field, primed for generating circularly polarized attosecond pulses~\cite{medisauskas_generating_2016,hickstein_non-collinear_2015}. 

Importantly, a non-collinear arrangement allows the use of a single frequency for both beams. As an initial approximation, the superposition of two opposite circular polarizations creates, locally, a linearly polarized field which permits the generation of harmonics. Here, the relative phase between the beams changes as one moves transversally across the focus, and this rotates the direction of the local polarization of the driving fields, and that of the emitted harmonics with it. This forms a `polarization grating' for the harmonics, which translates into angularly separated circular polarizations in the far~field~\cite{hickstein_non-collinear_2015}.

Upon closer examination, however, the planes of polarization of the two beams are at a slight angle, which means that they have nonzero field components along the centerline of the system. At certain points these components will cancel, giving a linear polarization, but in general they will yield the elliptical polarization shown in \reffig{fig:figureoneb}. 

The possibility of forwards ellipticity in vacuum fields runs counter to our usual intuition, and so far it has only been considered in the context of a very tight laser focus~\cite{lin_tight-focus_2006}. Our configuration provides a flexible, readily available experimental setup. In particular, it allows the focal spot size (and therefore the laser intensity) to be decoupled from the degree of forwards ellipticity. 
This ability is crucial, since it allows the ionization fraction to be tuned for phase-matching (although reaching perfect phase matching conditions in the x-ray region requires very high pressure-length products because of the very long absorption lengths in the x-ray region).

To bring things on a more concrete footing, we consider the harmonics generated in a noble gas by two beams with opposite circular polarizations propagating in the $x,z$ plane (as in \reffig{fig:figureonea}) with wavevectors 
\begin{equation}
\mathbf k_\pm=k(\pm \sin(\theta),0,\cos(\theta)),
\end{equation}
where the angle $\theta$ to the centerline on the $z$ axis is typically small. The vector potential therefore reads
\begin{align}
&\vba  (\vbr,t)
=
\sum_\pm
\frac{F}{2\omega}
\begin{pmatrix}
\cos(\theta)\cos(\vbk_\pm\cdot\vbr-\omega t) \\
\pm\sin(\vbk_\pm\cdot\vbr-\omega t)\\
\pm\sin(\theta)\cos(\vbk_\pm\cdot\vbr-\omega t)
\end{pmatrix}
\nonumber \\ & \  =
\frac{F}{\omega}
\begin{pmatrix}
\phantom{-}  \cos(\theta)\cos(kz\cos(\theta)-\omega t)\cos(kx\sin(\theta)) \\
\phantom{-\sin(\theta)}  \cos(kz\cos(\theta)-\omega t)\sin(kx\sin(\theta))\\
-\sin(\theta)\sin(kz\cos(\theta)-\omega t)\sin(kx\sin(\theta))
\end{pmatrix}.
\label{full-field}
\end{align}

As an initial approximation, for small $\theta$, the polarization planes coincide, and the polarization becomes 
linear, with a direction which rotates across the focus: 
\begin{equation}
\vba(\vbr,t)
\approx
\frac{F}{\omega}
\begin{pmatrix}
\cos(kx\sin(\theta)) \\
\sin(kx\sin(\theta)) \\
0
\end{pmatrix}
\cos(\omega t),
\end{equation}
where we set $z=0$ and therefore just examine a single transverse plane. However, when taken in full, the vector potential has a slight ellipticity, with a maximal value of $\eps = \sin(\theta)$ when $kx\sin(\theta)=\tfrac{\pi}{2}$, in which case
\begin{equation}
\vba  (\vbr,t)
=
\frac{F}{\omega}
\begin{pmatrix}
0 \\
\:\:\cos(\omega t)\\
\sin(\theta)\sin(\omega t)
\end{pmatrix}.
\label{forwards-elliptical-field}
\end{equation}
This forwards ellipticity acts in the same direction as the magnetic Lorentz force of the beam, so it can be used to control its effects as well as measure it, as exemplified in~\reffig{fig:figureone}. As we shall show below, the field in~\eqref{forwards-elliptical-field} will produce even harmonics, through the symmetry breaking shown in~\reffig{fig:figureone}. Since the ellipticity of the full field~\eqref{full-field} varies across the focus, so does the strength of the even harmonics, and this spatial variation in their production is responsible for their appropriate far-field behaviour.

In experiments, the beam half-angle $\theta$ will typically be small, on the order of $1\si{\degree}$ to $5\si{\degree}$~\cite{hickstein_non-collinear_2015}, with corresponding ellipticities of up to $\eps=\sin(\theta)\approx 9\%$, which is enough to counteract even significant magnetic drifts while still maintaining a flexible experimental scheme. 

The generation of harmonics beyond the breakdown of the dipole approximation has been described in a fully-relativistic treatment~\cite{milosevic_relativistic_2002, milosevic_relativistic_2002-1}, but this can be relaxed to the usual Strong-Field Approximation~\cite{LewensteinHHG} with appropriate modifications to include non-dipole effects~\cite{walser_hhg_2000,kylstra_photon_2001, kylstra_photon_2002, chirila_nondipole_2002, chirila_analysis_2004}. 
If a single beam is present, non-dipole terms break the dipole selection rules and produce even harmonics, but these are polarized along the propagation direction and therefore do not propagate on axis. The use of multiple beams in the non-dipole regime allows for observable breakdowns of the selection rules~\cite{averbukh_stability_2002}, but the available results are only valid for restricted beam arrangements; here we extend the formalism of Kylstra et al.~\cite{kylstra_photon_2001, kylstra_photon_2002, chirila_nondipole_2002, chirila_analysis_2004} to arbitrary beam configurations.

We start with the Coulomb-gauge hamiltonian, with the spatial variation of $\mathbf A$ taken to first order in $\mathbf r$,
\begin{align}
\hat H_V & = \frac 12\left(\hat \vbp+\vba(\hat \vbr,t)\right)^2+\hat V_0
 \\ & = \frac 12\left(\hat \vbp+\vba(\vb 0,t)+(\hat \vbr\cdot\nabla)\vba(\vb0,t)\right)^2+\hat V_0.
\end{align}
(We use atomic units unless otherwise stated.) We then perform a unitary transformation to $\ket{\Psi_L}=e^{i\hat \vbr\cdot \vba(\vb0,t)} \ket{\Psi_V}$, as in the dipole case, and we define this as the length gauge. Here the hamiltonian reads
\begin{equation}
\hat H_L = \frac 12\left(\hat \vbp+(\hat \vbr\cdot\nabla)\vba(\vb0,t)\right)^2+\hat \vbr\cdot\vbf(t)+\hat V_0,
\end{equation}
with $\vbf(t)=-\tfrac{\partial\vba}{\partial t}(\vb0,t)$. Moreover, we neglect terms in $\left((\hat \vbr\cdot\nabla)\vba(\vb0,t)\right)^2$ for consistency, 
as they are of higher order in $kr$, to get our final hamiltonian
\begin{equation}
H_L = \frac{\hat{\vbp}^2}{2}+\hat \vbr\cdot\vbf(t)+\hat\vbr\cdot\gradA(t)\cdot\hat\vbp+\hat V_0
 = \hat H_\mathrm{las}+\hat V_0.
\label{length-gauge-hamiltonian}
\end{equation}
Here the gradient $\gradA(t)$ denotes a matrix whose $i,j$-th entry is $\frac{\partial A_j}{\partial x_i}(\vb0,t)$, so in 
component notation the laser-only hamiltonian reads
\begin{equation}
\hat H_\mathrm{las} = \frac{\hat\vbp^2 \!\! }{2}+\hat x_jF_j(t)+\hat x_j\frac{\partial A_k}{\partial x_j}(t)\hat p_k,
\label{laser-only-hamiltonian}
\end{equation}
with summations over repeated indices understood.

Calculating the harmonic emission caused by the hamiltonian \eqref{length-gauge-hamiltonian} is essentially as simple as in the dipole case, and one only needs to modify the continuum wavefunction to include the non-dipole term. The required states here are non-dipole, non-relativistic Volkov states, which obey the Schr\"odinger equation for the laser-only hamiltonian $\hat H_\mathrm{las}$ and which remain eigenstates of the momentum operator throughout. The dipole Volkov states are easily generalized by first phrasing them in the form
\begin{equation}
\volkov{\vbp}{t}=e^{-\frac i2 \int^t\vbpi(\vbp,\tau)^2\d \tau}\ket{\vbpi(\vbp,t)},
\label{volkov-state-definition}
\end{equation}
where $\ket{\vbpi(\vbp,t)}$ is a plane wave at the kinematic momentum $\vbpi(\vbp,t)=\vbp+\vba(t)$, and 
then finding appropriate modifications to $\vbpi(\vbp,t)$. It is then easy to show that, to first order in $1/c$, 
the non-dipole Schr\"odinger equation is satisfied if
\begin{equation}
\vbpi(\vbp,t)=\vbp+\vba(t)-\int^t\gradA(\tau) \cdot (\vbp+\vba(\tau))\d\tau,
\label{kinematic-momentum-nondipole}
\end{equation}
using the fact that $\int\gradA\d\tau \sim \frac{k}{\omega}A= \frac{1}{c}A$ for a monochromatic field.

The harmonic emission can then be calculated within the SFA using the scheme from~\citer{HHGTutorial} by using the non-dipole Volkov states as the continuum wavefunctions, which gives a harmonic dipole of the form
\begin{equation}
\vb D(t)
=
\int_\tn^t \!\!\d t'\!\!
\int\!\!\d\vbp\:
\vbd(\vbpi(\vbp,t))
e^{iS(\vbp,t,t')}
\vbf(t')\cdot\vbd(\vbpi(\vbp,t'))
+\cc
\label{initial-harmonic-dipole}
\end{equation}
with the action given by
\begin{equation}
S(\vbp,t,t')=I_p(t-t')+\frac12 \int_{t'}^t \vbpi(\vbp,\tau)^2\d\tau.
\label{initial-action}
\end{equation}
This is consistent with the results of Refs.~\citealp{kylstra_photon_2001, kylstra_photon_2002, chirila_nondipole_2002, chirila_analysis_2004}, and it is generally valid for single-beam settings.

However, in the presence of multiple beams one must modify the above formalism, because now the antiderivative 
\begin{equation}
\int^t\gradA(\tau) \cdot \vba(\tau)\d\tau,
\label{trouble-integral}
\end{equation}
from Eq.~\eqref{kinematic-momentum-nondipole}, can no longer be uniquely defined. In general, this occurs in the presence of multiple beams at nontrivial angles and with nontrivial relative phases, but when that happens the cross terms in $\gradA(\tau) \cdot \vba(\tau)$ are oscillatory about a nonzero average. This then causes the integral~\eqref{trouble-integral} to contain a linearly-increasing term.

\begin{figure}[hb]
\includegraphics[width=\columnwidth]{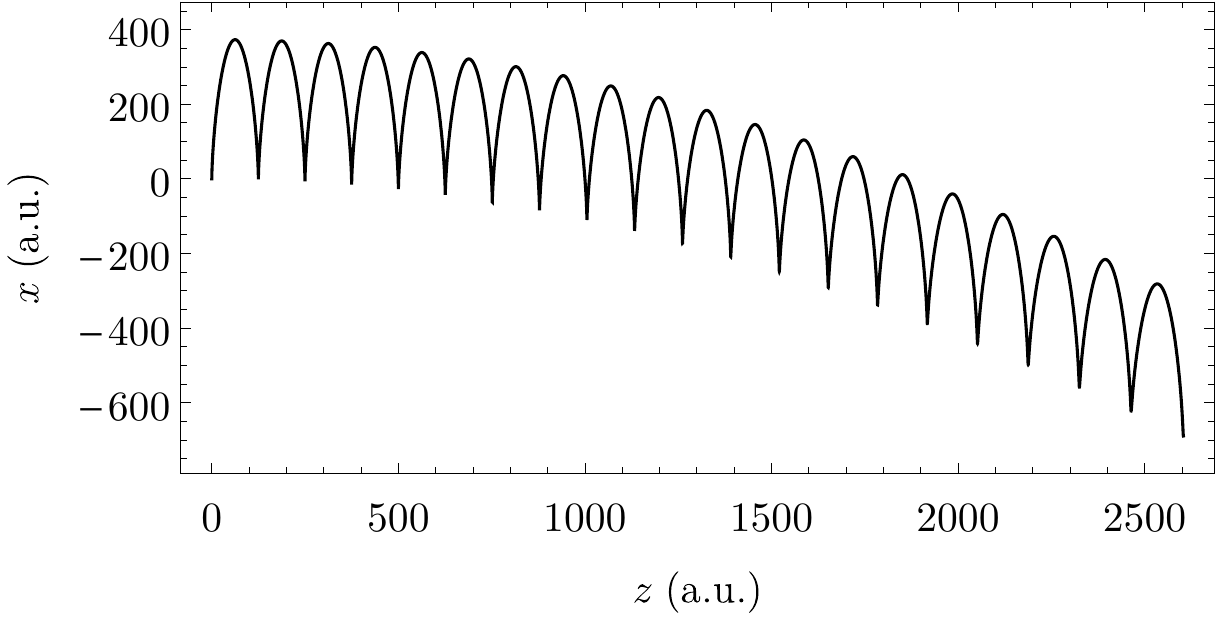} 
\caption{
The trajectory (shown here projected to the $x,z$ plane) of a particle released at rest at the peak of the field from~\eqref{full-field} exhibits oscillations about a central position which accelerates uniformly transversely across the focus, with a parabolic trajectory. The effect is generally slight, and the parameters here (\SI{1.6}{\micro\meter} beams at $\SI{e15}{\watt/\cm^2}$ with $\theta=20\si{\degree}$ at $kx\sin(\theta)=15\si{\degree}$) are somewhat exaggerated, but the loss of periodicity in \eqref{initial-harmonic-dipole} is serious. Similarly, the effect is only visually apparent over multiple oscillations, but even on the first oscillation the effect changes the time of recollision and therefore has a strong effect on the harmonic emission. This effect is also present for drivers with linear polarization in the common plane of propagation.
}
\label{fig:trajectory}
\end{figure}

This effect is real and physical, and it reflects the fact that the kinematic momentum $\vbpi$ is subject to a linear walk-off: that is, a constant force in the $x$ direction, orthogonal to the laser propagation direction, $z$, in addition to the usual oscillations (see~\reffig{fig:trajectory}). This constant force results from the interplay between the magnetic field and the $z$-direction velocity imparted by the elliptical electric field.

In practical terms, the effect is small but even in the first period it affects the timing of the ionization and recollision events, so it has a strong effect on the harmonic emission; as such, if not handled correctly it can introduce noise in a numerical spectrum at the same level as the signal.

From a mathematical perspective, this effect implies that states (\ref{volkov-state-definition},~\ref{kinematic-momentum-nondipole}) cease to be Floquet states of the laser hamiltonian when the dipole approximation breaks down. The Floquet states in this case are known in terms of Airy functions~\cite{Li-Reichl-Floquet} but those solutions are not particularly useful in this context. The nondipole Volkov states we use nevertheless form a basis of (approximate) solutions of the Schr\"odinger equation, but they now require an initial condition.

\newlength{\figurethreeheight}
\setlength{\figurethreeheight}{0.395 \columnwidth}
{
\setlength\tabcolsep{0mm}
\begin{figure*}
\scriptsize
\begin{tabular}{cccc}
  & (a) \SI{800}{nm}, \SI{e15}{W/cm^2}    &
    (b) \SI{1.6}{\micro\metre}, \SI{3.2e14}{W/cm^2} &
    (c) \SI{1.6}{\micro\metre}, \SI{e15}{W/cm^2}    \\
  \rotatebox{90}{\hspace{0mm} $|$dipole acceleration$|^2$ (arb.\,u.)} &
  \includegraphics[height=\figurethreeheight]{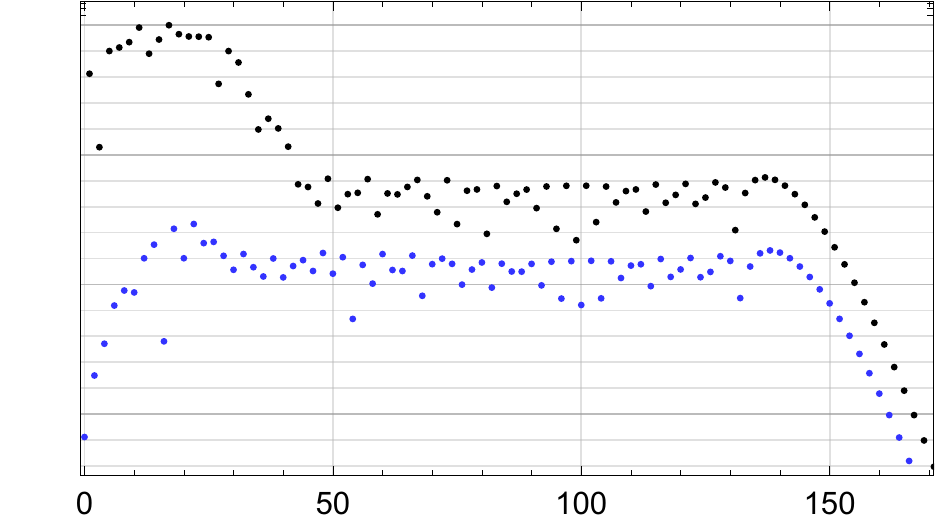} & 
  \includegraphics[height=\figurethreeheight]{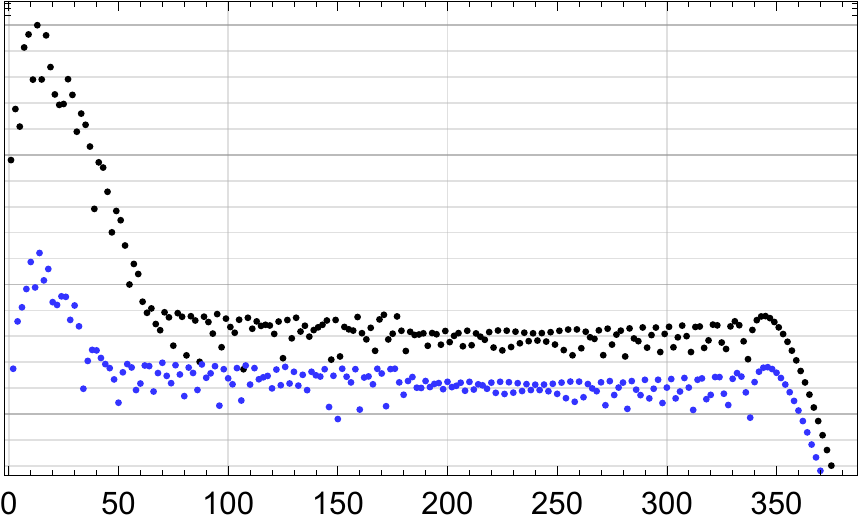} & 
  \includegraphics[height=\figurethreeheight]{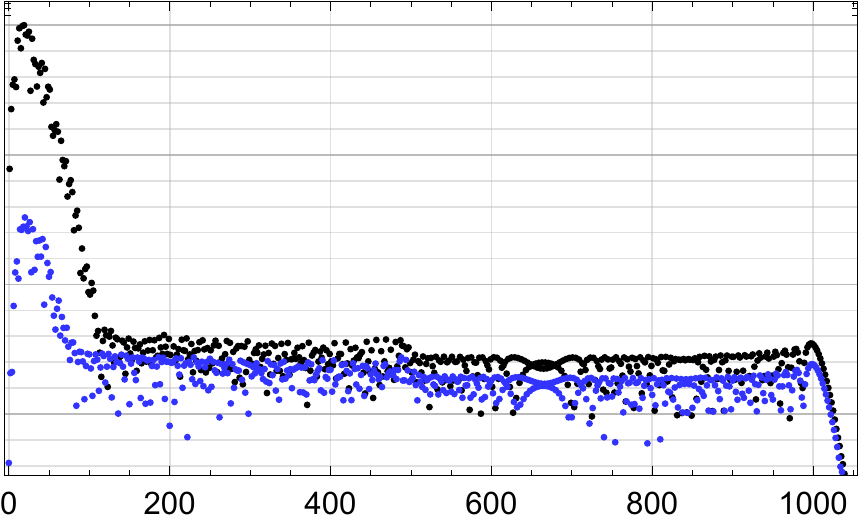} \\
  & harmonic order & harmonic order & harmonic order 
\end{tabular}

\caption{
Harmonic spectra ($\bullet$ odd and {\color{blue!80}$\bullet$} even harmonics) produced in helium in the non-dipole regime for intensities of $3.2\times 10^{14}$ and $\SI{e15}{W/cm^2}$ and monochromatic fields of \SI{800}{nm} and \SI{1.6}{\micro\metre} at a beam half-angle of $\theta=4\si{\degree}$, on arbitrary scales and eliminating $z$-polarized harmonics. The intensity ratio between even and odd harmonics varies from ${\sim}10^{-3}$ for \SI{800}{nm} drivers to ${\sim}10\%$ for strong mid-IR fields at $\SI{1.6}{\micro\metre}$ and $\SI{e15}{W/cm^2}$, which are still accessible with current technology. 
}
\label{fig:spectra}
{\color[gray]{0.8}\rule{0.9\linewidth}{0.3pt}}
\end{figure*}
}

{
To choose the appropriate initial condition, we note that the linear walk-off represents a secular term~\cite{Nayfeh_secular_terms} in these solutions, and we minimize the effect of this secular term by choosing an explicit reference time at the moment of ionization:
\begin{equation}
\vbpi(\vbp,t,t')=\vbp+\vba(t)-\int_{t'}^t \gradA(\tau) \cdot (\vbp+\vba(\tau))\d\tau.
\label{modified-kinematic-momentum-nondipole}
\end{equation}
This then trickles down to the action,
and similarly to the harmonic dipole.
}
 
\begin{figure}[hb]
\scriptsize
\begin{tabular}{cc}
  \rotatebox{90}{\hspace{3mm} $|$dipole acceleration$|^2$ (arb.\,u.)} &
  \includegraphics[width=0.9\columnwidth]{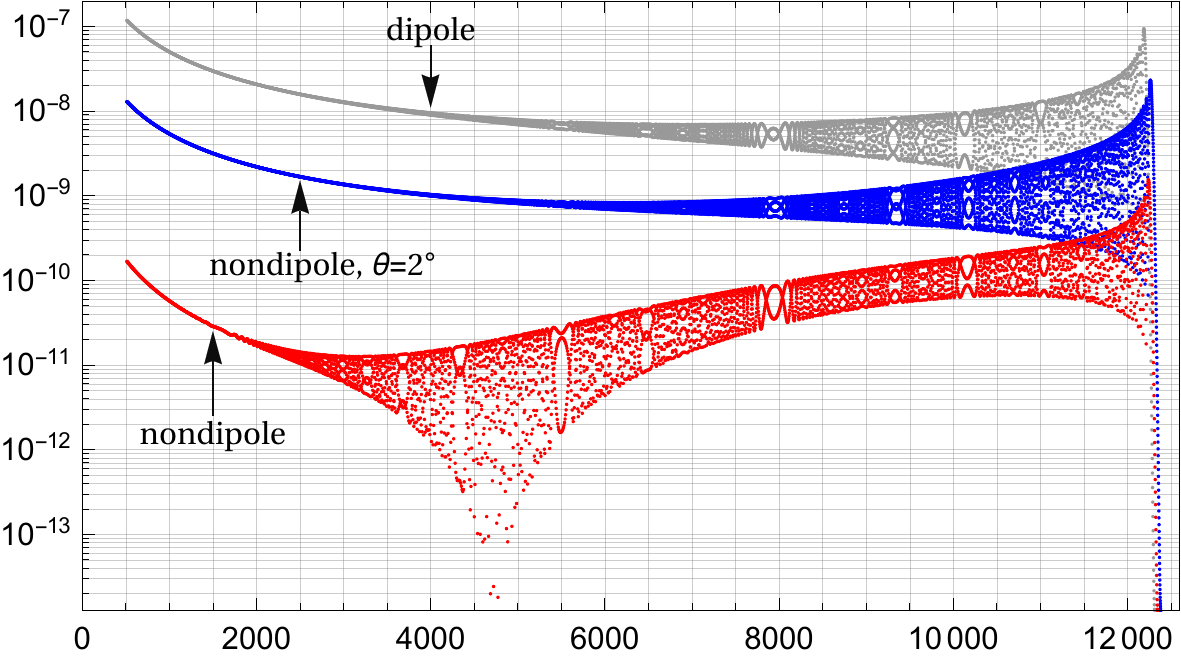} 
  \\ & harmonic order 
\end{tabular}
\caption{
  Above-threshold harmonic emission for a Ne$^{6+}$ ion in an $\SI{800}{nm}$ field of intensity $I=\SI{e17}{W/cm^2}$, calculated in the uniform approximation using the first pair of quantum orbits, and discarding $z$-polarized harmonics. For linear polarization, the dipole-approximation emission drops by two orders of magnitude when nondipole effects are included, but adding in even a small forwards ellipticity at $\theta=2\si{\degree}$ can help recover the harmonic emission. 
}
\label{fig:harmonic-recovery}
\end{figure}

This harmonic dipole is sufficient to evaluate the harmonic emission from arbitrary beam configurations, and it can be further simplified by the use of the saddle-point approximation for the momentum integral, and the uniform approximation \cite{figueira_uniform-approximation_2002,milosevic_long-quantum-orbits_2002} for the temporal integrations. Figs.~\ref{fig:spectra} and \ref{fig:harmonic-recovery} show our calculations of the single-atom response at locations in the focus where the forwards ellipticity is maximal. Our implementation is available from Refs.~\citealp{RB-SFA} and~\citealp{FigureMaker}.

At high intensities, shown in \reffig{fig:harmonic-recovery}, the presence of nondipole effects causes a drop-off in intensity~\cite{chirila_nondipole_2002}. Adding in a small amount of forwards ellipticity (at~$\theta=2\si{\degree}$) re-enables much of the harmonic emission, though further optimization is possible here.

The breaking of the intra-cycle symmetry is visible at much lower intensities, as shown in \reffig{fig:spectra} for fields at \SI{800}{nm} and \SI{1.6}{\micro\metre}. In particular, the non-dipole even harmonics begin to approach detectable intensities (between 0.1\% and 1\% of the intensity in the odd harmonics) even at \SI{800}{nm}, and they are on par with the odd harmonics at $\SI{1.6}{\micro\metre}$ and $\SI{e15}{W/cm^2}$. Such pulses can be produced using current optical parametric amplifiers, and they sit below the saturation intensity of helium, eliminating the need for highly charged species as a medium. The detection of non-dipole effects in HHG, then, can be done at rather moderate wavelengths and at intensities and with relatively simple experiments.

In addition to this, the even harmonics are also angularly separated from the dipole-allowed odd harmonics. This angular separation results from the conservation of momentum, and it has been clearly demonstrated for the dipole harmonics~\cite{hickstein_non-collinear_2015}: these must absorb an odd number of photons, but the conservation of spin angular momentum~\cite{fleischer_spin_2014, pisanty_spin-conservation_2014} requires the harmonic to form from $n$ photons of one beam and $n+1$ photons from the other, resulting in a net transverse momentum of $\pm\hbar k_x = \pm\hbar k\sin(\theta)$ for the odd harmonics. The even harmonics represent the parametric conversion of an even number of photons, via the tensor operator $\hat{\vbr}\otimes\hat{\vbp} \mathbin{:} \nabla \hspace{-0.15em} \vba$, and they can therefore absorb either zero transversal momentum (resulting in linear polarization along the $y$ axis) or $\pm2\hbar k_x$, with opposite circular polarizations. These even harmonics, then, appear at distinctly resolvable spots in the far field, which greatly simplifies their detection.

Finally, we note that it is the interferometric quality of our scheme that enables the detection of nondipole effects, by unbalancing (both in phase and in amplitude) the interferometer which would otherwise suppress the even harmonics, and this changes the scaling of this behaviour. In general, the wavepacket displacement scales~as
$d \propto {F^2}/{2c\omega^3}$,
and the wavepacket width goes as 
$\Delta x \propto {F^{1/2}}/\omega {(2I_p)^{1/4}} $~\cite{hatsagortsyan_laser_driven_2008},
so the normalized displacement scales as 
$$\zeta=\frac{d}{\Delta x}\propto \frac{(2I_p)^{1/4}F^{3/2}}{2c\omega^2}.$$
The strength of the even harmonics, which arises from an interferometric effect, is linear in $\zeta$, while the drift-induced reduction in harmonic emission follows the gaussian shape of the wavepacket and therefore scales with $\eta=\zeta^2=(d/\Delta x)^2$, which explains why the nondipole effects are accessible to current sources via our scheme but still some way in the future as regards the harmonic efficiency.

EP and MI acknowledge financial support from DFG and EPSRC/DSTL MURI grant EP/N018680/1. DDH, BRG, CGD, HCK and MMM thank AFOSR MURI grant FA9550-16-1-0121. EP thanks CONACyT and Imperial College London for support. BRG acknowledges support from the NNSA SSGF program.

\bibliographystyle{arthur} 
\bibliography{references}{}


\end{document}